# The Digital Evolution of Occupy Wall Street

**Michael D. Conover\*, Emilio Ferrara, Filippo Menczer, Alessandro Flammini**

Center for Complex Networks and Systems Research, School of Informatics and Computing, Indiana University, Bloomington, Indiana, United States of America

## Abstract

We examine the temporal evolution of digital communication activity relating to the American anti-capitalist movement Occupy Wall Street. Using a high-volume sample from the microblogging site Twitter, we investigate changes in Occupy participant engagement, interests, and social connectivity over a fifteen month period starting three months prior to the movement's first protest action. The results of this analysis indicate that, on Twitter, the Occupy movement tended to elicit participation from a set of highly interconnected users with pre-existing interests in domestic politics and foreign social movements. These users, while highly vocal in the months immediately following the birth of the movement, appear to have lost interest in Occupy related communication over the remainder of the study period.

**Funding:** The authors gratefully acknowledge support from the National Science Foundation (grant CCF-1101743), the Defense Advanced Research Projects Agency (DARPA) (grant W911NF-12-1-0037), and the McDonnell Foundation. The funders had no role in study design, data collection and analysis, decision to publish, or preparation of the manuscript.

**Competing Interests:** The authors have declared that no competing interests exist.

\* E-mail: midconov@indiana.edu

## Introduction

Information communications technologies play a crucial role in the development and persistence of many modern social movements [1–3]. Among these, the American anti-capitalist movement Occupy Wall Street ('Occupy') is remarkable for the prominent role social media, and in particular Twitter, played in facilitating communication among its participants [4,5]. Functioning as a high-visibility forum in which adherents and prospective participants could interact and share information, Twitter represented a valuable resource for supporting the movement's political and social objectives. In time, however, activity on the platform substantially diminished, mirroring the fading prominence of protest action on the ground. In light of this decline, we seek to understand more about the population from which Occupy drew its support, and specifically whether these individuals exhibited changes in behavior or social connectivity over the course of the movement's evolution.

The Twitter platform, like other information communication technologies, has the potential to confer a number of benefits to burgeoning social movements [6–8]. Chief among these is the opportunity to connect individuals in service of the dual goals of resource mobilization and collective framing [9]. These factors, well studied in the social sciences literature, are critical to the success of social movements. Resource mobilization refers to the process whereby a social movement works to marshal the physical and technological infrastructure, human resources, and financial capital necessary to sustain its ongoing activity [10,11]. Collective framing refers to the social processes whereby movement participants negotiate the shared language and narrative frames that help define the movement's identity and goals [12,13].

In related work [9], we report on evidence that Occupy users leveraged Twitter to communicate, at the local level, time-sensitive information about protest and police action. We also find that users relied on these channels to facilitate interstate communica-tion relating to the news media and narrative frames such as "We are the 99%," suggesting that long-distance communication on Twitter played a role in the collective framing processes that imbue social movements with a shared language, purpose and identity. This evidence indicates that Occupy participants used the Twitter platform to address critical issues facing any burgeoning social movement, and that during peak periods these streams were rich with actionable, relevant information.

To establish the extent of Occupy participant engagement with Twitter over time, here we study the total amount of Occupy-related traffic on the platform from September 2011 through September 2012. With respect to this measure of activity, we find that Occupy traffic has diminished by orders of magnitude relative to peak activity volumes in late 2011. This effect is evident even in concerted attempts to revive the movement's flagging levels of engagement, with activity returning to baseline within a week of May 1st, 2012 reoccupation efforts.

Finding little evidence of sustained activity, we turn our attention to Occupy participants themselves, in hopes of understanding how these users were changed as a result of engaging with the movement online. Using a random sample of 25,000 Occupy users, we study changes in behavior at the individual level with respect to attention allocation and social connectivity. From this analysis we are left to conclude that, on Twitter, Occupy evoked interest from a highly-interconnected community of users with pre-existing interest in domestic politics and foreign social movements. Though we find statistically significant changes in political interests and social connectivity over the study period, the magnitude of these changes pales in comparison to the amount of attention these individuals allocated to the Occupy Wall Street cause.

## Materials and Methods

### Twitter Platform

Twitter is a social networking platform that allows individuals to consume content from and contribute content to streams comprised of 140-character messages known as *tweets* [14]. The Twitter stream has been extensively explored in the recent literature, with focus on user activity modeling [15–19], content classification [20–23], sentiment analysis [24–26] and event detection [27–29]. Broadly speaking, there are two types of content streams: those associated with individual accounts and those associated with topic-specific tokens known as *hashtags*. By *following* one or more accounts, a user creates a personalized *feed* that aggregates into a single, private stream the content produced by the followed accounts. Hashtags, short tokens prepended with a pound sign (e.g., #taxes or #obama), allow the content produced by many individuals to be aggregated into a public, topic-specific stream including all the tweets containing a given token.

Although by default each user's tweets are publicly visible, the audience for an individual's content is largely limited to his or her network of immediate followers, attaining greater levels of visibility only when it is rebroadcast by large numbers of other users. By including a hashtag in a tweet, however, an individual can contribute content to a high-profile stream, and thereby engage with users who might never otherwise see the content. It is this kind of communication, which represents engagement with a topically cohesive community of users unconstrained by social network structure, that is the primary focus of this study.

In addition to engaging with different content streams, users can interact with one another in two primary ways. A user can *retweet* content produced by another individual, rebroadcasting it to his or her audience of followers, or *mention* another user in a tweet, which functions as a publicly-visible message targeting that individual.

### Data

We rely on two primary datasets extracted over a 15-month period from an approximately 5–10% sample of the entire public Twitter stream (https://dev.twitter.com/docs/streaming-apis/streams/public). In addition to information about the content and users associated with a tweet, the Twitter streaming API provides timestamp metadata that allow for the historical reconstruction of the time series presented in this study.

To identify Occupy-related content, we deem relevant any tweet containing a hashtag matching either #ows or #occupy*, where * represents a wildcard character. This set includes high-profile tags such as #occupy as well as location-specific tokens such as #occupyoakland and #occupyseattle. While this approach does not allow us to study content that does not contain an Occupy-specific hashtag, we argue that it is appropriate for two reasons. As outlined above, hashtags allow a user to reach an audience beyond his or her immediate followers, and it is this kind of expressly public engagement in which we are primarily interested. Moreover, while topic modeling techniques may allow for the analysis of untagged tweets, their use would introduce noise that could cloud the interpretation of any analytical results [30]. Based on the criteria outlined above, we produce a corpus of all sampled tweets containing at least one of these hashtags from the year-long period between September 1st, 2011 to August 31st, 2012. Referred to hereafter as the *Occupy corpus*, this dataset contains approximately 1.82 million tweets produced by 447,241 distinct accounts.

In addition to changes in activity explicitly related to the Occupy movement, we are also interested in changes to the behavior of individual users over time. To this end, we identified a random sample of 25,000 random users who produced at least one tweet in the Occupy corpus. We then produced a second corpus containing any tweet, regardless of content, produced by each account in this sample during the 15-month period spanning June 1st, 2011 through August 31st, 2012. Including tweets from the three-month period preceding the start of the Occupy Wall Street movement allows us to study the behavior of these users before, during, and after the movement's primary period of activity. Referred to hereafter as the *random sample*, this dataset contains approximately 7.74 million tweets produced by 25,000 unique users.

To facilitate analysis relating to the attention allocation habits of these individuals, we rely on three non-overlapping sets of hashtags: those related to Occupy Wall Street (defined above), a second set relating to foreign social movements, and a third relating to domestic political communication. As we are interested exclusively in the attention allocation habits of Occupy users, we identified the set of hashtags relating to domestic political communication and foreign social movements by manually inspecting the 300 hashtags most frequently used by individuals in the random sample. Table 1 lists the hashtags associated with each topic. While not exhaustive due to a long-tail use distribution, the 300 most popular hashtags account for 70.8% of all tagging activity, with the 300th most popular tag constituting just 0.027% of all tags. We therefore believe that the inclusion of additional tags in our topic lists is not likely to affect the results of this study.

### Methods

All of the analyses in this article rely on time series describing changes to measured quantities over the course of the study period. Each time series is produced by computing a single statistic on disjoint sets of tweets partitioned into adjacent, temporally non-overlapping bins of $k$ hours. For all of these analyses we use one of three temporal resolutions to reveal different characteristics of the signal under study: 12 hours, 24 hours, or one week.

At various times over the course of the study period, our system experienced service outages that affected our ability to collect data from the Twitter API. Amounting to 15 days in total, these periods are: September 29 to October 4, 2011; October 11–12, 2011; December 28–30, 2011; February 11–13, 2012; February 16–17, 2012; and May 28–31, 2012. Owing to the fact that the measures we employ reflect relative composition of the stream rather than its absolute volume, these outages do not unduly influence the statistical character of our results.

## Results

Let us first focus on the total number of tweets in the Occupy corpus over the course of the year. Figure 1 shows that, in general, Occupy traffic closely mirrors activity on the ground, and is characterized by peak levels during the month-long period following the movement's initial protests, with significantly diminished activity levels over the following eleven months. In terms of relative change, average levels of Occupy traffic in the second half of the period from September 17th, 2011 to August 31st, 2012 decreased 80.8% relative to the first half of the same period.

In light of this finding, we wish to gain insights into the character of the individuals from which Occupy drew its support. We begin by studying how Occupy user interests changed in time, examining the frequency with which 25,000 random individuals produced content relating to one of three topics: Occupy Wall Street, foreign social movements, and domestic politics. Based on the random sample described in §Data, the results of this analysis

**Table 1.** Lists of topic-specific hashtags.

| Domestic Politics | Social Movements |
| --- | --- |
| #tcot | #syria |
| #p2 | #bahrain |
| #teaparty | #egypt |
| #gop | #yemen |
| #anonymous | #libya |
| #obama | #tahrir |
| #tlot | #wiunion |
| #jobs | #iranelection |
| #ronpaul | #assange |
| #romney | #wikileaks |
| #sopa | #jan25 |
| #ndaa | #14feb |
| #obama2012 | #assad |
| #ocra | #greece |
| #twisters | #damascus |
| #sgp | #gaddafi |
| #politics | #feb14 |
| #solidarity | #scaf |
| #gop2012 | #antisec |
| #p21 | #arabspring |
| #topprog | #tunisia |
| #obamacare | #noscaf |
| #mapoli | #syrian |
| #acta | |
| #sotu | |
| #newt | |
| #santorum | |
| #mittromney | |
| #gopdebate | |
| #dem | |

Hashtags were manually selected from among the 300 most frequently used by individuals in the 25,000-person random sample of Occupy users.
doi:10.1371/journal.pone.0064679.t001

describe activity from June 1, 2011 to August 31, 2012, a period including the three months prior to the initial protest action.

As we are interested in the behavior of individuals who were active on Twitter at a given time, we identify the set of users $U_i$ from whom we observe at least one tweet at time step $i$, regardless of its content. Within this set we isolate, at each timestep, the set of users $U_{it}$ from whom we observe, in any of their tweets, at least one hashtag relating to topic $t$. The *engaged user ratio* $|U_{it}|/|U_i|$ describes the extent to which individuals chose to engage in communication relating to each of the three topic areas.

Among the set of users engaged with a topic, we next examine the extent to which that topic tends to dominate their content production activity. To accomplish this, let us consider, for each user $u \in U_{it}$, the collection $H_{iu}$ of hashtags contained in his or her tweets at time step $i$. From this we compute the proportion of each user's tagging activity that is associated with a given topic, $|H_{iut}|/|H_{iu}|$, where $H_{iut}$ is the set of tags from topic $t$ produced by $u$ at time step $i$. Averaging this value across all engaged users provides a lens on the behavior of these individuals as a whole, and is reported as the *engaged user attention ratio*. Figure 2 presents this value alongside the engaged user ratio to show how the amount of attention allocated to the three topics changed over time.

As expected, a large fraction of users produced Occupy related content during the period of peak activity, with more than 40% of sampled users allocating on average 64% of their attention to the topic during the third week following the initial protests. However, this intense focus on the subject is not sustained over the course of the following year, with the engaged user ratio decaying to less than 5% in the last three months of the study period. Moreover, comparing the engaged user attention ratios from the first half of the period following the initial Occupy protests ($\mu = .439$) to those from the second half ($\mu = .318$), we find that individuals who continue to produce Occupy content do so with significantly lower frequency. Computed using a two location t-test for a difference in sample means, we reject the null hypothesis ($p < 10^{-3}$) that the mean of the engaged user attention ratios in the first half of the study period is greater than or equal to that of the observations in the second half of the study period, a finding suggestive of diminished enthusiasm even among the most persistent individuals.

With respect to foreign social movements and domestic political communication, we observe that users who would go on to engage

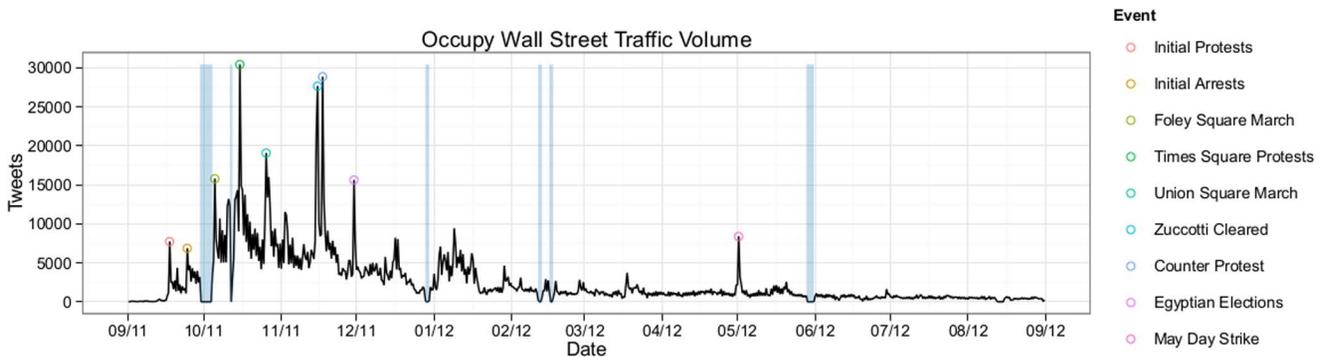

**Figure 1. Total number of tweets related to Occupy Wall Street between September 2011 and September 2012.** Each timestep represents a 12-hour period, with vertical blue bars overlaid on periods during which access to the Twitter streaming API was interrupted. Large bursts in activity tend to correspond to protest or police action on the ground, demarcated with circles. From left to right, the events are: initial Occupy Wall Street protest in Zuccotti Park; initial NYPD arrests of protesters; march from Foley Square to Zuccotti Park; protest at U.S. Armed Forces recruiting station in Times Square; protest in support of Iraq veteran injured by police-fired projectile; NYPD action to clear Zuccotti Park; protest against eviction from Zuccotti Park; first round of Egyptian elections; 'May Day' general strike and planned reoccupation of former encampments.
doi:10.1371/journal.pone.0064679.g001

   3   

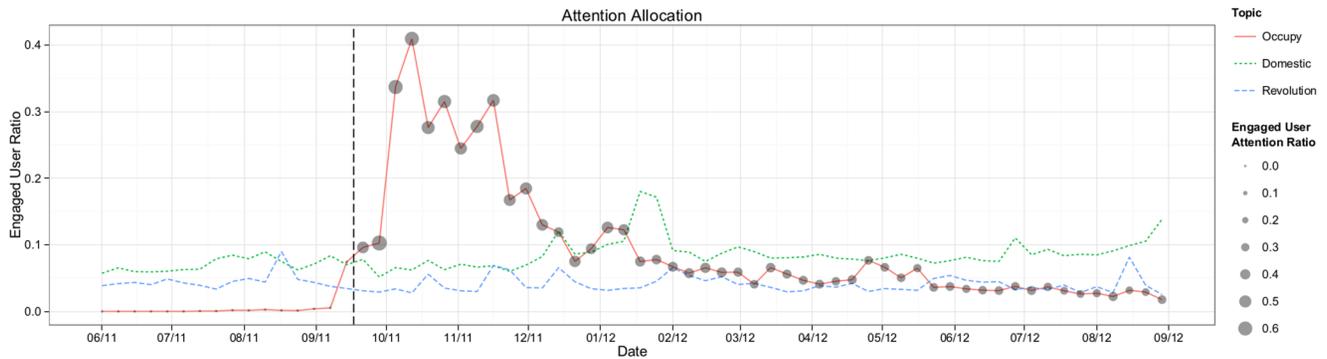

**Figure 2. Attention allocation of 25,000 randomly selected Occupy users to each of three topics: Occupy Wall Street, domestic politics, and revolutionary social movements.** Engaged User Ratio describes the proportion of active users in each timestep who produced at least one topically-relevant tweet. Engaged User Attention Ratio describes, among these users, the share of average attention allocated to each topic. The Engaged User Attention Ratio did not exhibit meaningful trends for either domestic politics or foreign social movements, and so it is omitted from the figure for sake of visual clarity. Refer to § Results for the full derivation of these measures. The dashed vertical line corresponds to the date of the first Occupy protest.
doi:10.1371/journal.pone.0064679.g002

with the Occupy movement online tended to exhibit interest in these topics before the initial protest activity in September, 2011. Comparing the engaged user ratios in the first 12 weeks of the study period with those observed during the last 12 weeks of the study period, we find a significant but small increase in domestic political communication activity. This conclusion is based on a two location t-test for a difference in sample means, in which we reject the null hypothesis ($p < 10^{-3}$) that the mean of the engaged user ratios from the first twelve weeks ($\mu = 0.066$) is greater than or equal to that of the latter half of the study period ($\mu = 0.077$). With respect to interest in foreign social movements, we observe a significant ($p < 0.05$) but small decrease in engagement for the same periods (from $\mu = 0.074$ to $\mu = 0.057$). These differences suggest that the changes in individual behaviors in response to the Occupy Wall Street movement were limited.

Finally, let us examine the extent to which Occupy users tended to interact with one another over the course of the study period. To this end we focus on the proportion of retweets and mentions produced by active users in the random sample that involved another user who produced at least one Occupy-related tweet during the year following the movement's inception. This proportion is computed with respect to all of a user's retweets and mentions, regardless of content, rather than just those related to Occupy Wall Street. Inspecting the 95% confidence interval bands in Figure 3 we observe a statistically significant increase in in-group retweet and mention activity during the peak period of Occupy activity, followed by a gradual decay to values approaching pre-Occupy levels. Comparing the fifteen week period before the inception of the movement to the fifteen week period at the study's close, we use a two location t-test to identify a small but significant increase in both in-group retweets ($p < 10^{-6}$) and mentions ($p < 10^{-3}$), with the mean connectivity increasing 5.1% for retweets and 3.2% for mentions. Although these changes are statistically significant, they can hardly be interpreted as evidence that this community's long-term social connectivity has been dramatically altered in response to participation in the Occupy Wall Street movement. Moreover, it's notable that even in the period preceding the Occupy events, nearly 30% of these individuals' targeted retweeting activity and almost a quarter of their mentioning were originated from or were directed to other Occupy users, suggesting that the movement elicited engagement from an already tightly interconnected community of users, rather than uniting disparate social groups behind a common cause.

## Discussion

While interest and activity relating to the Occupy movement has substantially diminished, one could envision that increased levels of engagement with the political process online might constitute a positive outcome for the movement's participants. Along these lines, however, Occupy users remain barely changed, exhibiting a slight increase in attention paid to domestic politics and a slight decrease in attention paid to foreign social movements. Relative to the dramatic behavioral changes these users exhibited in the early stages of the movement, and the magnitude of Occupy-related communication in general, these changes constitute a somewhat underwhelming long-term effect.

Similarly, a supporter of the movement might take as a promising outcome increased levels of interaction among Occupy users. Such a scenario could indicate that these individuals formed a more tight-knit community over the course of the year, creating social and communication bonds that may help to facilitate the efficient spread of information, potentially even reinforcing individual propensity for offline activity [31,32]. The data, however, provide little evidence to indicate that Occupy precipitated a dramatic rewiring of these users' information sharing networks. While we observe significant increases in the proportion of in-group retweet and mention activity during the movement's peak, the trend suggests that these values are slowly returning to those observed before the movement's birth. What's more, in the months preceding the initial protests we find evidence indicating that these users were already highly interconnected, with more than a quarter of their directed communication (either retweeting or mentioning) involving another individual who would go on to create Occupy related content.

Taken together, these data suggest that, on Twitter, the Occupy movement tended to elicit participation from a set of highly interconnected users with pre-existing interests in domestic politics and foreign social movements. These same users, while highly vocal in the months immediately following the movement's birth, appear to have lost interest in Occupy-related communication over the remainder of the study period, and have exhibited only marginal changes in their attention allocation habits and social connectivity as a result of their participation.

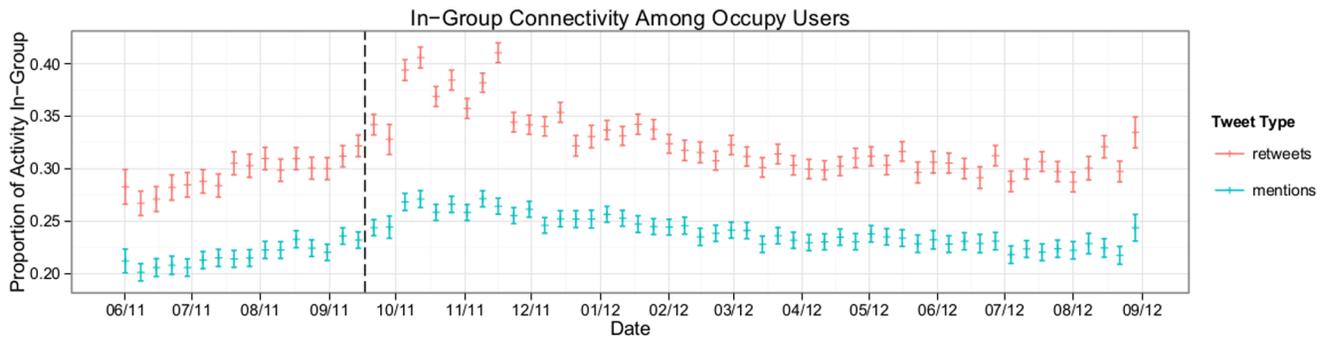

**Figure 3. Proportion of all retweet and mention traffic, regardless of content, from 25,000 randomly selected Occupy users involving another individual who produced at least one Occupy-related tweet.** Shown are means and 95% confidence intervals for each time step. The dashed vertical line corresponds to the date of the first Occupy protest.
doi:10.1371/journal.pone.0064679.g003

These findings should not be taken to suggest that the Occupy movement itself has failed, as an argument can be made that the movement played a role in increasing the prominence of social and economic inequality in the public discourse. Though it would be unreasonable to argue that users could have maintained the frenetic pace of Occupy's earliest days, it is doubtless that supporters may have hoped for a more sustained discourse than is evident from the near-complete abandonment of these once high-profile communication channels.

## Acknowledgments

Thanks to Twitter for making data available through their streaming API; and to Karissa McKelvey, Clayton Davis, Bruno Gonçalves, Jacob Ratkiewicz, and other current and past members of the Truthy Project at Indiana University (cnets.indiana.edu/groups/nan/truthy) for facilitating access to this data.

## Author Contributions

Conceived and designed the experiments: MDC EF FM AF. Performed the experiments: MDC EF. Analyzed the data: MDC EF FM AF. Wrote the paper: MDC EF FM AF.